\input harvmac
\input amssym.def
\input amssym.tex
\noblackbox
\newif\ifdraft

\catcode`\@=11
\newif\iffrontpage
\newif\ifxxx
\xxxtrue

\newif\ifad
\adtrue
\adfalse

\noblackbox

\parindent0pt

\def\{{\lbrace}
\def\}{\rbrace}

\def\R{{\Bbb R}}

\def\a{\alpha}
\def\b{\beta}

\def\s{\sigma}
\def\g{{g}}

\def\p{\partial}

\def\gz{\mathrel{\mathop g^{\scriptscriptstyle{(0)}}}} 
\def\go{\mathrel{\mathop g^{\scriptscriptstyle{(1)}}}} 
\def\gt{\mathrel{\mathop g^{\scriptscriptstyle{(2)}}}} 
\def\ao{\mathrel{\mathop a^{\scriptscriptstyle{(1)}}}} 

\def\Box#1{\mathop{\mkern0.5\thinmuskip
           \vbox{\hrule\hbox{\vrule\hskip#1\vrule height#1 width 0pt\vrule}
           \hrule}\mkern0.5\thinmuskip}}
\def\box{\displaystyle{\Box{7pt}}}
%
%


%
\def\abstract#1{
\vskip .5in\vfil\centerline
{\bf Abstract}\penalty1000
{{\smallskip\ifx\answ\bigans\leftskip 2pc \rightskip 2pc
\else\leftskip 5pc \rightskip 5pc\fi
\noindent\abstractfont \baselineskip=12pt
{#1} \smallskip}}
\penalty-1000}
\def\tr{{\rm tr}}

\vbadness=10000

\parindent10pt

%
\lref\AGMOO{O.~Aharony, S.S.~Gubser, J.~Maldacena, H.~Ooguri and Y.~Oz,
``Large N field theories, string theory and gravity,''
hep-th/9905111.}
\lref\DS{S.~Deser and A.~Schwimmer,
``Geometric classification of conformal anomalies in arbitrary dimensions,''
Phys.\ Lett.\ {B {\bf 309}}, 279 (1993)
[hep-th/9302047].}
\lref\FG{C. Fefferman and R. Graham, ``Conformal Invariants'',
Ast\`erisque, hors s\'erie, 1995, p.95.}
\lref\wittenone{E.~Witten,
``Anti-de Sitter space and holography,''
Adv.\ Theor.\ Math.\ Phys.\ {2} (1998) 253,
hep-th/9802150.}
\lref\Pen{R.~Penrose and W.~Rindler, ``Spinors and Spacetime'',
vol 2, CUP 1986, chapter 9.}
\lref\APTY{O.~Aharony, J.~Pawelczyk, S.~Theisen and S.~Yankielowicz,
``A note on anomalies in the AdS/CFT correspondence,''
Phys.\ Rev.\ {D60} (1999) 066001,
hep-th/9901134.}
\lref\HS{M.~Henningson and K.~Skenderis,
``The holographic Weyl anomaly,''
JHEP {\bf 07} (1998) 023,
hep-th/9806087.}
\lref\BPB{L.~Bonora, P.~Pasti and M.~Bregola, ``Weyl cocycles'',
Class. Quantum. Grav. 3 (1986) 635.}
\lref\CFL{A.~Cappelli, D.~Friedan and J.I.~Latorre,
``C theorem and spectral representation,''
Nucl.\ Phys.\ {B352} (1991) 616.}
\lref\BH{J.D.~Brown and M.~Henneaux,
``Central Charges In The Canonical Realization Of Asymptotic Symmetries:
An Example From Three-Dimensional Gravity,''
Commun.\ Math.\ Phys.\ {104} (1986) 207.}
\lref\Strings{
http://strings99.aei-potsdam.mpg.de/cgi-bin/viewit.cgi?speaker=Theisen}
\lref\Cardy{J.L.~Cardy,
``Is There A C Theorem In Four-Dimensions?,''
Phys.\ Lett.\ {\bf B215} (1988) 749.}
\lref\BGNNO{S.~Nojiri and S.D.~Odintsov,
``On the conformal anomaly from higher derivative gravity in
AdS/CFT  correspondence,'' hep-th/9903033;
M.~Blau, K.S.~Narain and E.~Gava,
``On subleading contributions to the AdS/CFT trace anomaly,''
JHEP {\bf 09} (1999) 018
hep-th/9904179.}
\lref\ISTY{C.~Imbimbo, A.~Schwimmer, S.~Theisen and S.~Yankielowicz,
``Diffeomorphisms and holographic anomalies,''
Class.\ Quant.\ Grav.\  {\bf 17}, 1129 (2000)
[hep-th/9910267].}
\lref\MALDA{J.~M.~Maldacena,
``The Large N limit of superconformal field theories and supergravity,''
Adv.\ Theor.\ Math.\ Phys.\  {\bf 2}, 231 (1998)
[hep-th/9711200].}
\lref\ST{A.~Schwimmer and S.~Theisen,
``Entanglement Entropy, Trace Anomalies and Holography,''
Nucl.\ Phys.\ B {\bf 801}, 1 (2008)
[arXiv:0802.1017 [hep-th]].}
\lref\DESC{B.~Zumino, Y.~-S.~Wu and A.~Zee,
``Chiral Anomalies, Higher Dimensions, and Differential Geometry,''
Nucl.\ Phys.\ B {\bf 239}, 477 (1984).}
\lref\DILATON{A.~Schwimmer and S.~Theisen,
``Spontaneous Breaking of Conformal Invariance and Trace Anomaly Matching,''
Nucl.\ Phys.\ B {\bf 847}, 590 (2011)
[arXiv:1011.0696 [hep-th]].}
\lref\KS{
Z.~Komargodski and A.~Schwimmer,
``On Renormalization Group Flows in Four Dimensions,''
JHEP {\bf 1112}, 099 (2011)
[arXiv:1107.3987 [hep-th]];\hfill\break
Z.~Komargodski,
``The Constraints of Conformal Symmetry on RG Flows,''
JHEP {\bf 1207}, 069 (2012)
[arXiv:1112.4538 [hep-th]].}
\lref\WITTEN{J.~Wess and B.~Zumino,
``Consequences of anomalous Ward identities,''
Phys.\ Lett.\ B {\bf 37}, 95 (1971);
E.~Witten,
``Global Aspects of Current Algebra,''
Nucl.\ Phys.\ B {\bf 223}, 422 (1983).}
\lref\Elvang{H.~Elvang, D.~Z.~Freedman, L.~-Y.~Hung, M.~Kiermaier,
R.~C.~Myers and S.~Theisen,
``On renormalization group flows and the a-theorem in 6d,''
JHEP {\bf 1210}, 011 (2012).
[arXiv:1205.3994 [hep-th]].}
\lref\Schwarz{J.~H.~Schwarz,
``Highly Effective Actions,''
[arXiv:1311.0305 [hep-th]].}
\lref\Manohar{
I.~Low and A.~V.~Manohar,
``Spontaneously broken space-time symmetries and Goldstone's theorem,''
Phys.\ Rev.\ Lett.\  {\bf 88}, 101602 (2002)
[hep-th/0110285].}
\lref\Ogievetsky{
E.~A.~Ivanov and V.~I.~Ogievetsky,
``The Inverse Higgs Phenomenon in Nonlinear Realizations,''
Teor.\ Mat.\ Fiz.\  {\bf 25}, 164 (1975).}
\lref\KuzenkoOne{
S.~M.~Kuzenko and I.~N.~McArthur,
``Quantum metamorphosis of conformal symmetry in N=4 super Yang-Mills theory,''
Nucl.\ Phys.\ B {\bf 640}, 78 (2002)
[hep-th/0203236].}
\lref\Kallosh{
R.~Kallosh, J.~Kumar and A.~Rajaraman,
``Special conformal symmetry of world volume actions,''
Phys.\ Rev.\ D {\bf 57}, 6452 (1998)
[hep-th/9712073].}
\lref\Jevicki{
A.~Jevicki, Y.~Kazama and T.~Yoneya,
``Generalized conformal symmetry in D-brane matrix models,''
Phys.\ Rev.\ D {\bf 59}, 066001 (1999)
[hep-th/9810146].
}
\lref\HenningsonTwo{
M.~Henningson and K.~Skenderis,
``Holography and the Weyl anomaly,''
Fortsch.\ Phys.\  {\bf 48}, 125 (2000)
[hep-th/9812032].}
\lref\Manes{
J.~Manes, R.~Stora and B.~Zumino,
``Algebraic Study of Chiral Anomalies,''
Commun.\ Math.\ Phys.\  {\bf 102}, 157 (1985).}
\lref\Bellucci{
S.~Bellucci, E.~Ivanov and S.~Krivonos,
``AdS / CFT equivalence transformation,''
Phys.\ Rev.\ D {\bf 66}, 086001 (2002), [Erratum-ibid.\ D {\bf 67}, 049901 (2003)]
[hep-th/0206126].}
\lref\McArthur{
I.~N.~McArthur,
``Nonlinear realizations of symmetries and unphysical Goldstone bosons,''
JHEP {\bf 1011}, 140 (2010)
[arXiv:1009.3696 [hep-th]].}
\lref\Max{
T.~Andrade, M.~Ba\~nados and F.~Rojas,
``Variational Methods in AdS/CFT,''
Phys.\ Rev.\ D {\bf 75}, 065013 (2007)
[hep-th/0612150].}
\lref\DeserYX{
S.~Deser and A.~Schwimmer,
``Geometric classification of conformal anomalies in arbitrary dimensions,''
Phys.\ Lett.\ B {\bf 309}, 279 (1993)
[hep-th/9302047].}
\lref\AK{
O.~Aharony and Z.~Komargodski,
``The Effective Theory of Long Strings,''
JHEP {\bf 1305}, 118 (2013)
[arXiv:1302.6257 [hep-th]].
}

\Title{\vbox{
\rightline{\vbox{\baselineskip12pt}}}}
{Comments on the Algebraic Properties
of Dilaton Actions
\footnote{$^{\scriptscriptstyle*}$}{\sevenrm
Partially  Supported by the Center for Basic Interactions of the Israeli Academy of Sciences.
A.S. also acknowledges support from the Alexander von Humboldt-Foundation.
}}
\vskip 0.3cm
\centerline{A. Schwimmer$^a$ and S. Theisen$^b$}
\vskip 0.6cm
\centerline{$^a$\it  Weizmann Institute of Science, Rehovot
76100, Israel}
\vskip.2cm
\centerline{$^b$ \it Max-Planck-Institut f\"ur Gravitationsphysik,
Albert-Einstein-Institut, 14476 Golm, Germany}
\vskip0.0cm

%

\abstract{
We study the relation between the dilaton action and sigma models
for the Goldstone bosons of the spontaneous breaking of the conformal group.
We argue that the relation requires that the sigma model is diffeomorphism invariant.
The origin of the WZW terms for the dilaton is clarified and  it is
shown that  in this approach the dilaton WZW term is necessarily accompanied
by a Weyl invariant term  proposed before from holographic considerations.}
\Date{\vbox{\hbox{\sl {}}
}}
\goodbreak

%


\newsec{Introduction}

Chiral anomalies are well understood algebraically \DESC\Manes.
Their general form can be obtained by considering
the theory in $d=2n$ as a boundary of a $d+1$ dimensional manifold. The action in $d+1$
dimensions is the local
Chern-Simons action and since this action is gauge invariant only up
to a boundary term the correct anomaly
is reproduced by this boundary term. In a sense this could be considered
as a manifestation of ``holography".

When the chiral symmetry is spontaneously broken Goldstone bosons are present.
Since the Goldstone bosons have to reproduce the chiral anomalies, specific features
of their action  (``sigma-model" in the following) follow from the
aforementioned structure:
besides a local term in $d=2n$ dimensions which realizes nonlinearly the symmetry
there is a second term \WITTEN\ (``WZW term" in the following) which lives in $d+1$
dimensions and reproduces the anomaly through the above mechanism.

In the present note we want to study in detail the analogous problems for trace anomalies.
Following the explicit calculation of the trace anomalies in the AdS/CFT
duality \HS\ it was realized \ST\ that a mechanism rather analogous
to the one described above for chiral anomalies is at work:
the gravitational action in $d+1$ dimension plays the role of the Chern-Simons
action and a particular subgroup of the $d+1$-dimensional diffeomorphism
acts as the analogue of the gauge transformations producing the anomalies at the boundary.

When the conformal symmetry is spontaneously broken the Goldstone boson
(in the following ``the dilaton") should reproduce the trace anomalies \DILATON\KS.
The effective action with this property can be constructed and a WZW term appears.
Compared with the general properties of the chiral Goldstone bosons action
outlined above the dilaton action has strange features: the WZW term is local directly in
$d=2n$ and does not seem to have any higher dimensional origin.

In order to understand this feature we rely on the basic
distinguishing property of the dilaton:
even though the spontaneous breaking of the conformal symmetry in  Euclidean signature
is the breaking of the $SO(d+1,1)$ group to $SO(d)\times T_d$
there is only one Goldstone boson,
the dilaton \Manohar\Ogievetsky\McArthur.
This is of course a consequence of the fact that all the
conformal currents can be
constructed in terms of the energy momentum tensor,
their conservation being the consequence
of tracelessness. The gauging of the $SO(d+1,1)$ currents is replaced by diffeomorphism
invariance and by the Weyl symmetry.

In the broken phase one would start naively  with a sigma-model on the
$SO(d+1,1)/[SO(d)\times T_d]$ coset.
We propose that the reduction from the $d+1$ fields parametrizing the
coset to the single dilaton is achieved
by a special new, characteristic feature of the sigma-model: diffeomorphism
invariance in $d=2n$ dimensions
for the invariant term and in $d+1$ dimensions for the WZW term, respectively.
By choosing a particular parametrization the coset coordinates are reduced
to a single field and the WZW
term becomes explicitly local.

We will formulate the sigma-model in a general metric background.
Since the input $d$-dimensional metric  should give rise in the sigma-model
to a metric depending
on $d+1$ coordinates -- the Goldstone boson fields -- we are led from
the beginning to consider a ``holographic" setup. Moreover,
the metric in the sigma-model action should admit the action of
a group isomorphic to the Weyl group which makes the connection to
holography even stronger. In spite of that  the freedom for dilaton actions
constrained by the  algebraic approach is much larger than the one
which follows from a strict application of holography. In particular,
as we will discuss in detail, there is no relation between the
$d+1$ dimensional actions and solutions
we are using in the construction.

Applying the above mentioned  procedure both for the invariant terms of the
dilaton action and the one reproducing the trace anomalies (the ``WZW" part)
we get an  interesting connection between the two once one imposes the condition
that there is no potential for the dilaton, a necessary condition for the
spontaneous breaking of conformal invariance. Our conclusion is that the
special action proposed for the dilaton in a holographic setup in
\MALDA\ has a general algebraic origin being normalized by the ``$a$" trace anomaly
whenever conformal invariance is spontaneously broken.

The paper is organized as follows:

In Section 2 we construct the $d$ dimensional part of the
reparametrization invariant sigma-model
and we show how it reduces to the Weyl invariant part of the dilaton action.
In Section 3 we construct the $d+1$ dimensional reparametrization invariant WZW term
and we reduce it to the dilaton WZW term. We discuss the relation between invariant terms
and the WZW terms following form the requirement of vanishing potential for the dilaton.
The relations between WZW terms  corresponding
to different even dimensions is made explicit.
In the last section we discuss various applications of the formalism developed and
possible generalizations.
In Appendix A we review the holographic calculations of trace anomalies and the
realization of Weyl symmetry in holography which motivate the choices of the
explicit metric backgrounds in Sections 2 and 3.

Related and complementary discussions of some of the aspects addressed here can be found
in \Kallosh\Jevicki\KuzenkoOne\Bellucci\Schwarz.
While these references rely on supersymmetry,
this is not assumed here.

\newsec{The Weyl invariant part of the dilaton sigma-model}

We will follow here an algebraic approach though, as we will see,
the results have an immediate holographic interpretation.

Consider in $d$ dimensions the breaking of the conformal
group $SO(d+1,1)$ to
the Poincar\'e group $ SO(d)\times T_d $. The coset of Goldstone bosons can be parametrized
by $d+1$ fields $X^{\mu}(x^i)$ where $ \mu=1,...,d+1$ and $i=1,...,d$.
The metric on the space of the $X$ fields is $AdS_{d+1}$ with isometry $SO(d+1,1)$ such that
the broken isometries are nonlinearly realized.

Since we want to construct the analogue of the  ``gauged sigma-model" we allow a more
general metric  $G_{\mu\nu}(X^{\tau})$ on which the Weyl transformations act.
The condition this metric has to fulfill in order to serve our purposes are:

\noindent
a) It should be a $d+1$ metric but a functional of a $d$ dimensional
metric $g_{ij} , i,j=1,...,d$
\eqn\basic{G_{\mu\nu}=F_{\mu\nu}[g_{ij}]\,.}
\noindent
b) It should admit the action of a group isomorphic to the Weyl transformations such that
\eqn\basicc{\delta _{\sigma} G= F[g_{ij} \exp{2\sigma(x)}]-F[g_{ij}]}
where $\delta _{\sigma}$ denotes the action on the $d+1$ dimensional metric isomorphic
to the Weyl transformations.

\noindent
c) For $g_{ij}=\delta_{ij}$ it should reduce to the natural metric on the
${SO(d+1,1)}\over{SO(d)\times T_d}$ coset which is $AdS_{d+1}$.

A class of metrics which satisfy the above requirements are solutions of
$d+1$ dimensional ``bulk" actions which admit $AdS_{d+1}$ solutions,
specified by the boundary metric $g_{ij}$ (which in the continuation we will
denote by $g_{ij}^{(0)}$) in the Fefferman-Graham gauge.
Obviously this class satisfies the above requirements, the group action isomorphic
to Weyl transformations
being the PBH transformations as explained in the Appendix.

The connection to holography is now obvious though we stress that we will use only
the algebraic properties of the solution. In particular the specific action to
which the metric is a solution will not play a role. It is an interesting question
if there are metrics which satisfy the above requirements not arriving from a holographic
construction.

The natural building blocks  for the gauged sigma-model are  the induced metric:
\eqn\indu{h_{ij}(x)=G_{\mu\nu}(X^{\tau}(x))\partial_i X^{\mu}(x) \partial_j X^{\nu}(x)}
and the second fundamental form.
While for an ordinary sigma-model we would take as an action e.g. $\delta^{ij}  h_{ij}$,
here we insist on reparametrization invariance in $d$ dimensions which, together
with the field redefinition invariance
present in \indu, will allow us
to project to the dilaton. Therefore the minimal  sigma-model action
having these properties is:
\eqn\acty{S={1\over\ell^d}\int d^d x \sqrt{\det h_{ij}}\,.}
If from the beginning we choose $G_{\mu\nu}$ to be in the FG gauge,
we split the fields $X^{\mu}$ into $X^i(x)$ $i=1,...,d$ and $\Phi(x)$
in the $``\rho"$-direction.
Now we can achieve the reduction to the dilaton action by choosing the gauge:
 \eqn\gauge{ X^i(x)=x^i}
and the gauge fixed action becomes:
\eqn\mald{S={1\over\ell^d}\int d^d x {\sqrt{\det g_{ij}(x,\Phi(x))}
\over{\Phi^{d/2}(x)}} \sqrt{1+{\ell^2{g^{ij}(x,\Phi(x))\partial_i\Phi\partial_j\Phi(x)}
\over{4\Phi(x)}}}\,.}
For the particular case of the AdS metric  we recognize the expression proposed in
\MALDA\ as representing the action for the displacement of the  brane which
breaks the gauge and conformal symmetry on the  $N=4$ Super Yang-Mills Coulomb branch.

In order to exploit the symmetries of the action it is convenient to start with
its unfixed form \acty.
Since in the FG gauge the metric is determined by its boundary
value $g^{(0)}_{ij}(X)$, the action is a functional of $g^{(0)}_{ij}$ and $X^{\mu}$.
The action  has  symmetries of two kinds:

a) Field transformations of the $X^{\mu}$ fields relating
two different background metrics $G_{\mu\nu}$.
These transformation make explicit the variation under a change of $g^{(0)}_{ij}$.
The transformations are inherited from residual gauge transformations in the FG gauge,
i.e. the PBH transformation parametrized by $\sigma(X^j)$ and $X^j$-dependent field
transformations parametrized by $\zeta^{i}(X^j)$. We rewrite the PBH transformations
of the Appendix, making it explicit that in the framework of the
sigma-model we deal with field transformations at fixed coordinates $x^i$:
\eqn\diffeo{
 \Phi'= \Phi(1+2\sigma(X^j))}
\eqn\diffeoo{X'^i =X^i-a^i(X^j,\Phi(x))-\zeta^{i}(X^j)}
where
\eqn\ais{
a^i(X,\Phi(x))={\ell^2\over2}\int_0^{\Phi(x)} d\rho'g^{ij}(x,\rho')\partial_j\sigma(X)\,.}
Such a transformation changes $g^{(0)}_{ij}$ by:
\eqn\cha{\delta g^{(0)}_{ij}(X^k) =2 \sigma(X^k) g^{(0)}_{ij} (X^k)
+\nabla_{i}\zeta_j(X^k) +\nabla_j\zeta_i(X^k)}
where the covariant derivatives are constructed with $g^{(0)}_{ij}$
and all the functional dependences are on $X^j$.

b) Reparametrizations of the $x^i$ variables parametrized by $ \xi^{i}(x^k)$:
\eqn\rep{ x'^i=x^i - \xi^{i} (x^k)}
under which the ``fields" $X^{\mu}$ transform as:
\eqn\fiel{ \delta X^{\mu}(x^k)= \xi ^{i}\partial_{i} X^{\mu}(x^k)\,.}
We are now ready to study the symmetries of the ``projected" action
\mald\ in the special coordinates \gauge,
the action being now a functional just of $g^{(0)}_{ij}$ and $\Phi$.

After the transformation \diffeo\ the special choice
\gauge\ is not anymore respected.  In order to reinstate it\foot{For a
related discussion see \Elvang.}
we should make
a reparametrization with the special choice of the
parameters $\xi^i$:
\eqn\choice{ \xi^i(x^k) =a^{i}( X^k=x^k,\Phi(x)) +\zeta^{i}(X^k=x^k) }
Therefore the action \mald\ will be  invariant under a joint transformation:
\eqn\traa{\delta g^{(0)}_{ij}(x)= 2 \sigma(x) +\nabla_i \zeta_j +\nabla_j \zeta_i}
and
\eqn\trab{\delta \Phi(x)= 2 \sigma (x) \Phi(x)
+ [ a^{i}( X^k=x^k,\Phi(x)) +\zeta^{i}(X^k=x^k) ] \partial_{i} \Phi(x)}
where we indicated the places where the field $X^k$ was replaced with the
variables $x^k$ using the gauge \gauge.

Equations \traa, \trab\ make explicit the invariance of \mald\
under the Weyl transformations.
The transformations are non-anomalous since there is no boundary
term which could be the source
of violation in the above classical argument.

Equation \trab\ shows that in order to have the usual transformation of the
dilaton field $\tau$  under Weyl transformation:
\eqn\dila{\tau \rightarrow \tau+\sigma}
there is a field redefinition relating $\Phi$ and $\tau$.
Using  \trab\ the field redefinition
can be found iteratively giving an expansion in powers of $ e^{2\tau}$ where the
terms of ${\cal O}(e^{2 n\tau})$ contain $2(n-1)$ derivatives.\foot{The discussion here
generalizes that of \Elvang\ to a curved metric.}
The $0$-th order
solution of \trab\ is:
\eqn\Phizero{\Phi^{(0)}= e^{2 \tau}\,. }
For the higher order terms in the
iterative solution we make the most general Ansatz and require \trab\ to be
satisfied. (For simplicity and without loss of generality, we set $\zeta^i=0$.)
To proceed we need the $\Phi$-expansion  of $a^i$ which depends on the
higher order terms in the FG expansion of the metric. Both can be found
in \ISTY.
In this way we find
\eqn\Phione{{1\over \ell^2}\Phi^{(1)}={1\over2}e^{4\tau}(\nabla\tau)^2
+\a\, e^{2\tau}\hat R}
and
\eqn\Phitwo{\eqalign{&{1\over \ell^4}\Phi^{(2)}
=e^{6\tau}\left({R^{ij}\nabla_i\tau\nabla_j\tau\over 4(d-2)}
-{R (\nabla\tau)^2\over8(d-1)(d-2)}
+{1\over4}\nabla^i\tau\nabla^j\tau\,\nabla_i\nabla_j\tau+{7\over16}(\nabla\tau)^4\right)\cr
\noalign{\vskip.2cm}
&\quad+\a\, e^{2\tau}\left({1\over2}\hat\nabla^i\tau\hat\nabla_i\hat R
+\hat\nabla^i\tau\hat\nabla_i\tau\,\hat R\right)
+e^{2\tau}\left(\beta_1\hat{\box}\hat R+\beta_2\hat R^2
+\beta_3\hat R^{ij}\hat R_{ij}+\beta_4\hat C^{ijkl}\hat C_{ijkl}\right)}}
Here $\a$ parametrizes the homogeneous solution of \trab\ at ${\cal O}(\p^2)$ and
$\b_1,\dots,\b_4$ parameterizes the homogeneous solutions at ${\cal O}(\p^4)$.
The hatted quantities are built from the invariant metric $\hat g_{ij}=e^{-2\tau}g_{ij}$
and $g_{ij}=g^{(0)}_{ij}$.

Plugging   this into \mald\ we obtain the ``universal minimal"
invariant part of the action (cf. Appendix A):
\eqn\geneuniv{ S= {1 \over \ell^d} \int d^dx \sqrt{\hat g}
\left(1+{\ell^2\over 4(d-1)}\hat R
+{\ell^4\over32(d-2)(d-3)}(\hat E_4-\hat C^2)+{\cal O}(\p^6)\right)}
%
%
Additional terms can be obtained by adding to \acty\
reparametrization invariant terms depending on curvatures constructed
from $h_{ij}$ as well as from the second fundamental form.\foot{Such terms were 
considered, in a somewhat different context, in \AK.}  
Also, we have set $\a=\b_i=0$ (cf. \Phione,\Phitwo).
Note that in $d=2$ the ${\cal O}(\p^2)$ terms are ${1\over 4}\int\sqrt{g}R$ while in
$d=4$ the ${\cal O}(\p^4)$ terms are ${1\over 64}\int\sqrt{g}(E_4-C^2)$.

If one keeps the homogeneous terms in $\Phi$, one finds the additional terms
\eqn\Deltainv{
\eqalign{
&\Delta S={1\over\ell^d}\int d^d x\sqrt{\hat g}\Bigg\lbrace
-{d\,\ell^2\over2}\a\hat R+{\ell^4\over 8}d(d+2)\a\hat R^2
+{\ell^4\over8}{(d-2)\over(d-1)}\a^2\hat R^2\cr
&\qquad\qquad\qquad\qquad\qquad-{d\,\ell^4\over2}\Big(\b_1\hat{\box}\hat R
+\b_2\hat R^2+\b_3\hat R^{ij}\hat R_{ij}+\b_4 \hat C_{ijkl}\hat C^{ijkl}\Big)
\Bigg\rbrace\,.}}
The above analysis made for the minimal action can be repeated after
adding terms depending on curvatures built with the induced metric
and/or the second fundamental form. From the analysis above it is clear
that while the final form of the action  will depend on these terms the field redefinition
of $\Phi$ in terms of $\tau$ is universal.
A unique feature of the minimal action which we will use in the following is that it
is the only term written in terms of the $\Phi$ field which contains
a ``potential" of the dilaton field $\tau$. In flat space this is simply
${1\over \Phi^{d/2}}$, while after the field redefinition it gives rises
to the $\sqrt{\hat g}$ term.

Another application of our formalism is the study of the symmetries
of \mald\ for the special case
$g^{(0)}_{ij}=\delta_{ij}$, i.e. AdS background metric.
The AdS  metric is invariant under special
conformal transformations accompanied by an appropriate
Weyl transformation, i.e. in the notation of \diffeo:
\eqn\spec{\zeta^i= {1 \over 2} \epsilon^i\, x^2 -x^i (\epsilon\cdot x)}
with $\epsilon^i=$ const. and
\eqn\specon{ \sigma(x)=-{1\over d} (\p\cdot\zeta)= \epsilon\cdot x\,. }
For a flat background the $a^i$ have a very simple form:
\eqn\sspec{a_i={\ell^2 \over 2 }\Phi(x)\partial_{i} \sigma(x)
={\ell^2\over 2}\epsilon_i\Phi(x)\,.}
Therefore \mald\ for a flat $g^{(0)}_{ij}$ is invariant
under a reparametrization transformation:
\eqn\repa{x'^i=x^i-{1 \over 2} \epsilon^i x^2 +x^i (\epsilon\cdot x)
-{\ell^2 \over 2}\epsilon^i \Phi(x)}
followed by a field transformation:
\eqn\fie{\Phi'(x')=\Phi(x) +2 \epsilon\cdot x\, \Phi(x)}
which is the defining symmetry of \mald\ in \MALDA.

\newsec{The WZW term }

We want to construct a sigma-model term which reproduces the trace anomalies.
This should be an action on
a $d+1$ dimensional manifold with boundary and should share with the
invariant term the property of reducing
to a functional just of the dilaton field and the boundary metric.
We will achieve that by requiring that the action is invariant under
reparametrizations in $d+1$ dimensions. We define
\eqn\ind{f_{\alpha\beta}=G_{\mu\nu}(X)\,\partial_{\alpha}X^{\mu}\partial_{\beta} X^{\nu}}
where $G_{\mu\nu}$ is the bulk metric in FG gauge and
$\alpha,\beta=1,...,d+1$. The fields $X^{\mu}$ depend now on
$d+1$ coordinates  which we split from the beginning
into $x^i$, $i=1,...,d$ and $\rho$. The boundary of the manifold is
at $\rho=0$. The WZW action is then
\eqn\WZWact{S_{\rm WZW}={1\over\ell^d}\int d^{d}x\,d\rho\, \sqrt{\det f_{\alpha\beta}}\,.}

We remark that the classical solution metric which is used in \ind\
should not be necessarily a solution
of a minimal action like \WZWact: only its transformation properties enter our arguments.

Since the embedding has the same dimension as the space all the relevant information
is in the boundary conditions of the embedding fields.
We split $X^\mu$ into $X^i(x,\rho)$ and $\Phi(x,\rho)$.
The symmetries of the action are again:

a) Field transformations relating backgrounds defined by different $g^{(0)}_{ij}$.
These transformations, which involve fields at fixed coordinates,
will have exactly the same form as in the previous section, i.e. \diffeo\ and \ais.

b) Reparametrizations in  $d+1$ dimensions.

\noindent
We choose again a special set of coordinates by:
\eqn\speccc{ X^i(x,\rho)=x^i\,.}
Then the action \WZWact\ becomes:
\eqn\WZWfix{S_{\rm WZW}={1\over2\,\ell^d} \int d^d x\, d\rho\,
\partial_{\rho} \Phi {{\sqrt{\det g_{ij}(x,\Phi(x,\rho))}} \over\Phi(x,\rho)^{1+d/2}}\,.}
We see that a change of variable between $\rho$ and $\Phi(x,\rho)$ is possible in the
integral at each fixed $x$  such that the action \WZWfix\ depends
on $\Phi(x,\rho)$ just through  its boundary value $\Phi(x,\rho=0)$.

The symmetry transformations which leave the gauge condition \speccc\ unchanged are again
PBH transformations accompanied by reparametrizations:
\eqn\traa{\delta g^{(0)}_{ij}(x)= 2 \sigma(x) +\nabla_i \zeta_j(x) +\nabla_j \zeta_i(x)}
and
\eqn\trab{\delta \Phi(x,\rho)= 2 \sigma (x) \Phi(x,\rho)
+ [ a^{i}( X^k=x^k,\Phi(x,\rho)) +\zeta^{i}(X^k=x^k) ] \partial_{i} \Phi(x,\rho)}
$\sigma(x)$ and $\zeta^i(x)$  being the parameters of the transformation.
We remark that  as in the previous section $a^i$ are determined by the
PBH transformations such that
$\Phi$ is treated just as an expansion  parameter, its $x$-dependence not being acted upon.

At the boundary $\rho=0$,  $\Phi(x,\rho=0)$ has exactly the same
Weyl transformation property as $\Phi(x)$
in the previous section  and the field redefinition relating it
to the dilaton field $\tau(x)$ is identical.

As long as $\Phi(x,0)$ and $g^{(0)}_{ij}$ are transformed simultaneously
the action will be invariant.
This is the analogue in this set up of the fact that whatever the
generating functional of a CFT $W(g)$ is,
if we define $\hat g_{ij}= e^{-2\tau} g_{ij}$ then $W(\hat g)$  will be invariant under a
joint transformation of the metric and the dilaton.
The expression which is local and reproduces
the anomalies i.e. the WZW term, is the difference $W(g)-W(\hat g)$.

From \trab\ it is clear that in flat space $\tau=0$ corresponds to $\Phi(x,\rho=0)=1$.
Therefore, using the change of variable
from $\rho$ to $\Phi(x,\rho)$ we define the WZW action by
\eqn\WZWfin{S_{\rm WZW}={1\over2\,\ell^d}\int\limits^{\Phi(x,0)}\limits_{1}\!\!\!\! d\Phi\,
d^dx { {\sqrt{\det g_{ij}(x,\Phi)}}\over \Phi^{1+d/2}}\,.}
We expect that the expression \WZWfin\ has a well defined limit for $ d \rightarrow 2n$.

We discuss now the way \WZWfin\ reproduces  the trace anomalies. Under \traa,\trab\
the metric $\g^{(0)}_{ij}$ transforms by a Weyl transformation.
If this transformation is accompanied
by the appropriate transformation of $\Phi$ or equivalently of $\tau$,
this will be an invariance.
Therefore the variation will not get a contribution from the upper limit of integration.
On the other hand for the lower limit of integration the expression would be invariant
if  the lower limit which corresponds to $ \Phi(x,0)=1$
transformed as follows from \trab,
i.e. by the amount
\eqn\var{ \delta(\Phi=1)= 2 \sigma(x)\,.}
Since the lower  limit is kept fixed the variation is the compensating contribution
coming from the integrand multiplied with \var:
\eqn\ano{\delta S_{\rm WZW}=- {1\over2\,\ell^d} \int d^d x\, 2 \sigma(x)
\sqrt{\det  g_{ij}(x,\Phi=1)}\,.}
As discussed in the Appendix, in the expression \ano\ the anomalies in $d=2n$ are
to be found among the terms with no $\ell$ dependence. With this identification
it is clear that the type A anomalies \DS\
following from \ano\ are the ones calculated a long time ago \HS,\ISTY.

One can now explicitly work out the WZW part in the dilaton action in an
external metric following from \WZWfin. Using the FG expansions for $g_{ij}(x,\rho)$
in \HS\ and \ISTY\ we find in $d=4$:
\eqn\dilfourd{\eqalign{&S_{\rm WZW}=\int\!\!\sqrt{g}\,
d^4 x\Bigg( {1\over 4\,\ell^4}\Big(1-e^{-4\tau}\Big)
-{1\over 24\, \ell^2}R\Big(1-e^{-2\tau}\Big)
+{1\over 4\,\ell^2}(\p\tau)^2 e^{-2\tau}\cr
&+{1\over 64}\bigg(\tau\,E_4
+4\Big(R^{ij}-{1\over2}R g^{ij}\Big)\nabla_i\tau\nabla_j\tau
-4(\nabla\tau)^2\,\box\tau+2(\nabla\tau)^4\bigg)
+\Big({\gamma\over2}-{1\over 64}\Big)\,\tau\, C^2
\,.}}
Here $\gamma$ is a combination of the two parameters which specify the homogeneous
solution of the PBH transformation for $g^{(2)}$ \ISTY. They could be fixed by
explicitly solving the equations of motion of the bulk gravitational action. For the
Einstein-Hilbert action with cosmological constant, which is dual do ${\cal N}$
SYM theory, $\gamma=0$ and one finds that the two anomaly coefficients $a$ and $c$ are
the same.
In deriving \dilfourd\ we have set to zero the free parameters
which appear in the solution of \trab,
i.e the coefficients $\a$ and $\b_i$ in \Phione\ and \Phitwo.

An alternative way of writing \dilfourd\ is
\eqn\dilfourdalt{\eqalign{&S_{\rm WZW}=\int\!\!d^4 x\Bigg\lbrace
-{1\over 4\ell^2}\left(\sqrt{\hat g}-\sqrt{g}\right)
+{1\over 24 \ell^2}\left(\sqrt{\hat g}\hat R-\sqrt{g}R\right)\cr
&+{1\over 64}\bigg(\tau\,E_4
+4\Big(R^{ij}-{1\over2}R g^{ij}\Big)\nabla_i\tau\nabla_j\tau
-4(\nabla\tau)^2\,\box\tau+2(\nabla\tau)^4\bigg)
+\Big({\gamma\over2}-{1\over 64}\Big)\,\tau\, C^2
\Bigg\rbrace\,.}}

The expression \dilfourdalt\ above shows clearly that besides the standard
terms of the dilaton action which reproduce the trace anomaly there are
invariant terms including the ``dilaton potential" term $\sqrt{\hat g}$
whose normalization is completely
fixed by the ``$a$" anomaly.  In Appendix A this fact is analysed in detail showing
that it has an algebraic origin and it is independent
both on the metric used and a possible replacement of the ``minimal action"
\WZWfin\ by an action having additional curvature terms.

As shown in the previous section the action \mald\ is the only one among
the invariant actions which has a potential term for the dilaton.
Since the total effective action for  the dilaton cannot have a potential term
for the dilaton this term has to be cancelled by the one appearing
in the WZW part \dilfourdalt. We conclude therefore that \mald\
with a normalization $``-a"$ should necessarily accompany \dilfourdalt. This unexpected
conclusion is a consequence of our embedding the dilaton in the space of the $d+1$
Goldstone bosons with the symmetry structure assumed.
In the normalization of the type A anomaly coefficient s.t.
$\delta_\s W=a\int\s E_4+{\rm type\, B}$, the total dilaton action is
\eqn\total{
16\,a \big(S+4\,S_{\rm WZW}\big)\,.}
In this combination the dilaton potential cancels and the anomaly is reproduced
with the correct normalization. The same relative factor was derived in \MALDA\
from the requirement of supersymmetry in the form of a no-force requirement
for a probe D3-brane in $AdS_5$.

Alternative arguments
for the special role of \mald\ were recently put forward in \Schwarz.

If we want to calculate the WZW terms $S^{(0)}_{\rm WZW}$,
i.e. the dilaton self-interaction terms in flat external metric  from \WZWfin\
with $g^{(0)}_{ij}=\delta_{ij}$, we obtain a deceptively simple looking
universal expression:
\eqn\WZWterm{S^{(0)}_{\rm  WZW}=-{1\over d\,\ell^d}\int d^d x
\left({1\over\Phi(x,0)^{d/2}}-1\right)\,.}
For each dimension we should look at the term independent on $\ell$.
In principle this calculation should check the relative normalizations
of the terms in different dimensions
following from the general relation between type A anomalies \ISTY,\ST.

\newsec{Discussion.}

The special feature we used for constructing the sigma-model actions which reduced to
the dilaton action was their diffeomorphism invariance.
Using this invariance we could gauge away
some of the ``would be Goldstone bosons" ending with  the relevant one,
the dilaton. We believe
that this is a general feature whenever a space-time  symmetry is
spontaneously broken. It would
be interesting to have a general treatment of this pattern.\foot{This is 
similar in spirit to e.g. \Ogievetsky\McArthur, but in these references
backgrounds with isometries are considered while here no such 
assumption is made.} 

For the dilaton action the embedding gave new, unexpected  information,
i.e. that there is  a Weyl invariant
piece of the action, the ``minimal" one normalized by the $a$ anomaly
which should accompany the WZW term.
This feature appeared when we formulated  the actions in term of the $\Phi$ field,
the coordinate left unfixed after using diffeomorphism invariance.
The change of variable to the $\tau$ field, which realizes additively the
Weyl transformation, obscures
this connection. It is an open question if one can reformulate in
an invariant fashion the ``minimal" action in terms of the $\tau$ field.

We used extensively the ``gauging" of the sigma-model  in order to study
its symmetries. The gauging used was a coupling to a general
$d$ dimensional metric. We used  the fact
that the natural $AdS_{d+1}$ metric on the Goldstone boson space can
be related in the Poincar\'e patch to a $d$ dimensional flat metric.
Our ``gauging" was a natural deformation of this relation leading to
a natural appearance of the holographic set up. Alternative paths for coupling
the sigma-model to a metric, still respecting diffeomorphism invariance should be explored.
Once the above structure was picked we were led to consider metrics which were
solutions of equations of motion corresponding to a bulk action.
There is a large freedom  in the choice
of the bulk action and in particular there is no need that the action picked
for the sigma-model has any relation to the action which provided the metric solution.
In this sense the treatment seems to be purely algebraic. Nevertheless one should
investigate if this very relaxed holographic framework is really necessary
or the requirements we listed for the metrics can be achieved in a different way.

At a more basic level the gauging we used via the coupling to a metric might
not be the most natural one.  Following the analogy to the chiral
situation a coupling of the sigma-model to $SO(d+1,1)$ gauge fields would be more
natural. This would require  however the understanding of the relation (if any)
between  the $a$  Weyl anomaly and descent equations of the conformal group.
\bigskip

\noindent
{\bf Acknowledgements:} Useful discussions with O. Aharony and S. Yankielowicz
are gratefully acknowledged. We in particular thank Z. Komargodski for his 
collaboration at the early stages of this project and for his helpful 
comments throughout. 
S.T. thanks S. Kuzenko for the invitation, hospitality and discussions at
Western Australian University in Perth during the final stages.

\appendix{A}{Review of Trace Anomalies in Holography}

We will adopt the point of view that while AdS/CFT duality gives a physical
realization in which the trace anomalies of a CFT appear, one can abstract
from it general, algebraic properties
rather similar to the ``descent treatment" \DESC\ of chiral anomalies.

The basic setup involves  a $d+1$ (where $d=2n$) dimensional manifold  with the topology
of the Poincar\'e patch of AdS. For the $d$ dimensional boundary we will take a manifold
with Euclidean signature and the  topology of $\R^d$
(taking the topology to be e.g. $S^d$ is  unimportant as long as we discuss
local anomalies).

The metric $G_{\mu\nu}$ on the $d+1$ dimensional manifold can be brought to the
``Fefferman-Graham (FG in the following) gauge" ideally suited for the problem:
\eqn\ansatz{
ds^2=G_{\mu\nu} dX^\mu dX^\nu={\ell^2\over4}\left({d\rho\over\rho}\right)^2
+{1\over\rho}g_{ij}(x,\rho)dx^i dx^j\,.}
Here $\mu,\nu=1,\dots,d+1$ and $i,j=1,\dots,d$. The coordinates are
chosen such that $\rho=0$ corresponds to the boundary.
We will assume that $g_{ij}$ is regular at $\rho=0$.

We will consider $G_{\mu\nu}$, which are solutions of a gravitational equation of motion
determined through the FG expansion in terms
of the boundary value $g_{ij}^{(0)}(x)=g_{ij}(x,\rho=0)$.
The ``FG ambiguities" in the expansion determining $G_{\mu\nu}$
will not influence our calculations since
they appear one order higher than the expressions we use.

If we use as gravitational action the minimal one, i.e. Einstein term
and a negative cosmological constant,
then on the solution this action becomes:
\eqn\generalac{
S={1\over {\ell^d}}\int d^{d+1}X\sqrt{\det G_{\mu\nu}}\,.}
As we will discuss in the continuation, for the type A anomalies
related to the WZW term we are interested in,
\generalac\ gives the general expression up to an overall normalization.

The FG gauge is extremely useful since the symmetries playing an essential role
for trace anomalies appear as its residual gauge freedom \ISTY
\foot{Here our curvature
convention are the opposite to those of \ISTY, i.e.
$[\nabla_i,\nabla_j]V_k=R_{ijk}{}^l V_l$.}:

a) diffeomorphism transformations which act just on the $x$ variables, inducing
therefore  the ordinary $d$-dimensional
diffeomorphisms on the boundary

b) an additional subgroup of the $d+1$-dimensional diffeomorphisms depending on
one function $\sigma(x)$. The transformation (called PBH in the following) leaves
the metric in the FG gauge, i.e. ${\cal L}_\xi G_{\rho\rho}={\cal L}_\xi G_{\rho i}=0$,
with solution:
\eqn\PBH{\eqalign{
\rho'&=\rho\, e^{2\sigma(x)}\simeq \rho(1+2\sigma(x))\,,\cr
x'^i&=x^i-a^i(x,\rho)\,}}
where
\eqn\ais{
a^i(x,\rho)={\ell^2\over2}\int_0^\rho d\rho'g^{ij}(x,\rho')\partial_j\sigma(x)\,.}
Correspondingly, the metric $g_{ij}(x,\rho)$ transforms as:
\eqn\dg{
\delta g_{ij}(x,\rho)=2\sigma(1-\rho\partial_\rho)g_{ij}(x,\rho)
+\nabla_i a_j(x,\rho)+\nabla_j a_i(x,\rho)\,.}
The covariant derivatives are with respect to $g_{ij}(x,\rho)$
with $\rho$ fixed.

In particular the boundary value of the metric $g^{(0)}_{ij}$
transforms by an ordinary Weyl transformation:
\eqn\we{\delta g^{(0)}_{ij}(x)= 2 \sigma(x) g^{(0)}_{ij}(x)\,.}
Therefore the PBH transformation lifts the Weyl transformations into
an isomorphic subgroup of diffeomorphisms in $d+1$ dimensions.
The change between two metrics corresponding to boundary values related by \we\ can be
represented as a $d+1$ dimensional PBH transformation.

A diffeomorphism invariant action $S$ on a manifold with boundary transforms under a
diffeomorphism with parameters $\zeta^{\mu}$ as:
\eqn\act{\delta S =\int d^{d+1}X \partial_{\mu}(\zeta^{\mu} L)= 2 \sigma \rho L|_{\rho=0}}
where $L$ is the gravitational Lagrangian density and we used the
PBH diffeomorphism in the $\rho$-direction.
Therefore the anomalous Weyl variation of the action is expressed
directly through the boundary value
and  the trace anomalies can be read off directly from \act.

To make the discussion more concrete, we give the explicit expressions
needed for the calculation of anomalies in $d=4$.
The coefficients of the FG-expansion of the metric
\eqn\FGexpansion{
g_{i,j}(x,\rho)=\gz\!\!\!_{ij}(x)+\rho\go\!\!\!_{ij}(x)+\rho^2\gt\!\!\!_{ij}(x)+\dots}
are largely fixed
by PBH transformations \dg:
\eqn\gns{\eqalign{
\go_{ij}&=-{\ell^2\over d-2}\left(R_{ij}-{1\over 2(d-1)}R g_{ij}\right)\cr
\gt_{ij}&=c_1\,\ell^4\, C^2 g_{ij}+c_2\,\ell^4\, (C^2)_{ij}
+{\ell^4\over d-4}\Bigl\lbrace {1\over 8(d-1)}\nabla_i\nabla_j R
-{1\over 4(d-2)}\box R_{ij}\cr
&+{1\over 8(d-1)(d-2)}\box R\, g_{ij}-{1\over 2(d-2)}R^{kl}R_{ikjl}
+{d-4\over 2(d-2)^2}R_i{}^k R_{jk}\cr
&+{1\over (d-1)(d-2)^2}R R_{ij}+{1\over 4(d-2)^2}R^{kl}R_{kl}\,g_{ij}
-{3d\over 16(d-1)^2 (d-2)^2}R^2\,g_{ij}\Bigr\rbrace\,.}}
Here and below $\displaystyle{g=\gz}$ and likewise for
the covariant derivatives and the curvatures.
The coefficients $c_1$ and $c_2$ parametrize the homogeneous solutions of
\dg\ at ${\cal O}(\rho^2)$.
Inserting this into the expansion
\eqn\detg{
\sqrt{\det g(x,\rho)}=\sqrt{\det g}\left\lbrace
1+{1\over2}\rho\,\tr(\go)+\rho^2\Big[{1\over2}\tr\gt-{1\over4}\tr\big(\go{}\!^2\big)
+{1\over8}\big(\tr\go\big)^2\Big]+\dots\right\rbrace}
one finds
\eqn\trgexp{\eqalign{
\sqrt{\det g(x,\rho)}=\sqrt{\det g}\Bigg\lbrace 1-&{\ell^2\over 4(d-1)}\rho\,R
+\ell^4\,\rho^2\Big[\gamma\, C^2
+{1\over32(D-2)(d-3)}E_4\Big]+\dots\Bigg\rbrace}}
where PBH leaves $\gamma$ arbitrary\foot{E.g. for the minimal action
$\gamma=-{1\over32(d-2)(d-3)}$.}. $E_4$ is the dimensionally continued
four-dimensional Euler density $E_4=R_{ijkl}R^{ijkl}-4 R_{ij}R^{ij}+R^2$.
The contribution to the anomalies can be read off immediately from \trgexp.

We remark  that the type A anomaly is completely fixed by the PBH relation
modulo an over all normalization of the action.
The same normalization fixes in an unambiguous fashion the $\rho^0$ and $\rho^1$ terms.
This universality of the type A contribution with the accompanying two terms
is actually much more
general: any modification of the bulk action by higher curvature terms will lead
to the same grouping of the three terms, only modifying the overall normalization.
This is, in fact, easy to prove.

Consider an arbitrary (local) scalar $\Psi(x,\rho)$ built from the bulk curvature and its
covariant derivatives. It will have a FG expansion of the form
\eqn\FGPsi{
\Psi=\Psi^{(0)}+\rho\,\ell^2\,\Psi^{(1)}+\rho^2\,\ell^4\,\Psi^{(2)}+\dots}
The leading term is always a constant, namely $\Psi$ evaluated on AdS space.
Recalling that a term $\rho^n$ is accompanied by a curvature scalar
of the boundary metric with $2n$ derivatives, we see that necessarily
$\Psi^{(1)}\propto R$ and $\Psi^{(2)}$ must be a linear combination
of $\box R,\, R^2,\, R^{ij} R_{ij}$ and $C^2$.
Now we use the fact that $\Psi$ was a bulk scalar, i.e. under PBH it transforms
as
\eqn\deltaPsi{\eqalign{
\delta\Psi&=\xi^\mu\p_\mu\Psi=-2\,\sigma\,\rho\,\p_\rho\Psi+a^i\p_i\Psi\cr
&=-2\,\ell^2\,\rho\,\s\Psi^{(1)}+\rho^2\left(-4\,\ell^4\,\s\,\Psi^{(2)}
+\ell^2\!\ao{}\!\!^i\,\p_i\Psi^{(1)}\right)
+{\cal O}(\rho^3)}}
where $\p_i\Psi^{(0)}=0$ was used.
At ${\cal O}(\rho)$ this says that $\delta\Psi^{(1)}=-2\s\Psi^{(1)}$, but there
is no such curvature scalar. So we find at ${\cal O}(\rho^2)$ that
$\delta\Psi^{(2)}=-4\sigma\Psi^{(2)}$, from where we conclude that
$\Psi^{(2)}\propto C^2$ and the term proportional to
$E_4$ is unaffected.

This argument can be generalized to any even dimension.

Therefore the type A anomaly and the related $\sqrt{\hat g}$ term are a result
of the algebraic structure independent of the bulk action used. For convenience
we can use therefore  \generalac\ with an arbitrary overall normalization.
For the type A anomaly
in any even dimension  \generalac\ plays a role analogous to the Chern-Simons
action for chiral anomalies. The implications of the ``coupling" between the
type A anomaly and the $\sqrt{\hat g}$ term are  discussed in the main part of the paper.

The limiting process $\rho \rightarrow 0$  implicit in \act\ is not
entirely straightforward and we
make  now more precise some of its features. The actions of the
type of \generalac\ when evaluated
in the FG expansion have pole terms in $d-2n$ when treated in dimensional regularization.
In addition there are a finite number of terms with inverse powers of $\rho$
reflecting the
powerlike ultraviolet divergences present formally in the regularization which should be
however absent in the effective action of a CFT.
If the variation of the action is considered
as in \act\ the pole term in $d-2n$ is not anymore there and therefore the
limit $d\rightarrow 2n$ can be safely taken, however negative powers of $\rho$
are still present. Therefore before taking the limit in \act\ we should simply
discard these terms.\foot{Alternatively one can add explicit local
boundary terms to cancel them. For a clear discussion consult \Max.}

Alternatively we could use the following equivalent prescription
to isolate the trace anomalies:
we move the boundary to $\rho=1$ and we use \act. There are now in the
variation local terms
with positive and negative powers of the unique scale $\ell$ as coefficients.
The anomaly terms are simply the $\ell$ independent terms in the variation \act\
(modulo cohomologically trivial terms of course).

\vfill\eject

\listrefs

\bye